\documentclass{article}
\usepackage{amssymb}
\begin{document}
\title{Complete solutions of the Hamilton--Jacobi equation and the envelope method}

\author{G.F.\ Torres del Castillo \\ Departamento de F\'isica Matem\'atica, Instituto de Ciencias \\
Universidad Aut\'onoma de Puebla, 72570 Puebla, Pue., M\'exico \\[2ex]
G.S.\ Anaya Gonz\'alez \\ Facultad de Ciencias F\'isico Matem\'aticas \\ Universidad Aut\'onoma de Puebla, Apartado postal 165 \\ 72001 Puebla, Pue., M\'exico}

\maketitle

\begin{abstract}
It is shown that the parameters contained in any two complete solutions of the Hamilton--Jacobi equation, corresponding to a given Hamiltonian, are related by means of a time-independent canonical transformation and that, in some cases, a generating function of this transformation is given by the envelope of a family of surfaces defined by the difference of the two complete solutions. Conversely, in those cases, one of the complete solutions is given by the envelope of a family of surfaces defined by the sum of the other complete solution and the generating function of the canonical transformation. Some applications of these results to geometrical optics are also given.
\end{abstract}


\section{Introduction}
As is well known, in the framework of classical mechanics, the solution of the Hamilton equations can be obtained from a single complete solution of the Hamilton--Jacobi (HJ) equation (see, e.g., Ref.\ \cite{Ar}). What is no so widely known is that, as in the case of any first-order partial differential equation (PDE), there is an infinite number of complete solutions of the HJ equation that cannot be obtained from a complete solution substituting the parameters contained in it by functions of other parameters (see, e.g., Refs.\ \cite{CP} and \cite{IS}).

Lagrange found a method that leads, in principle, to the general solution of a first-order PDE starting from a complete solution (see, e.g., Ref.\ \cite{De} and the references cited therein). For instance, given a solution $z = f(x, y; a, b)$ of the first-order PDE in two variables
\begin{equation}
F \left( x, y, z, \frac{\partial z}{\partial x}, \frac{\partial z}{\partial y} \right) = 0, \label{fopde}
\end{equation}
containing two arbitrary parameters, $a$, $b$ (which, in this context, means that the solution is complete), substituting $b$ by some function of $a$, $b = \phi(a)$, one obtains the one-parameter family of solutions of (\ref{fopde}), $z = f \big( x, y; a, \phi(a) \big)$. Assuming that the equation
\begin{equation}
\frac{\partial f \big( x, y; a, \phi(a) \big)}{\partial a} = 0 \label{env1}
\end{equation}
can be solved for $a$ as a function of $x$ and $y$, $a = \chi(x, y)$, and eliminating $a$ one obtains a function
\begin{equation}
f \big( x, y; \chi(x, y), \phi(\chi(x, y)) \big)
\end{equation}
that is also a solution of (\ref{fopde}); this solution is the general solution of Eq.\ (\ref{fopde}) if $\phi$ is an arbitrary function, or another complete solution of Eq.\ (\ref{fopde}) if $\phi$ contains two new arbitrary parameters \cite{CP,IS}.

In the $xy$-plane, $f \big( x, y; a, \phi(a) \big) = 0$ represents a family of curves, parameterized by $a$, and $f \big( x, y; \chi(x, y), \phi(\chi(x, y)) \big)  = 0$ represents the {\em envelope}\/ of this family (roughly speaking, the envelope of a family of curves is a curve that touches tangentially each member of the family, see, e.g., Refs.\ \cite{CP} and \cite{IS}); therefore, this method of finding new solutions of a first-order PDE is equivalent to finding envelopes of families of curves (or surfaces, when there are more variables involved) that represent solutions of the equation.

One example of the use of the concept of envelope is encountered in geometrical optics where the propagation of the light can be described with the aid of wavefronts. According to the Huygens principle, each point of a wavefront is the source of secondary waves, and the new wavefronts are the envelopes of the secondary waves (see, e.g., Refs.\ \cite{Sy} and \cite{BW}). The wavefronts are the level surfaces of the eikonal function, which obeys a first-order PDE (the eikonal equation).

Actually, the concept of envelope is present in many places; for instance, finding the Legendre transform of a function of several variables amounts to finding the envelope of a family of surfaces.

The aim of this paper is to show that, for a given Hamiltonian, any two complete solutions of the HJ equation are related by means of a time-independent canonical transformation, and that one obtains, in a natural manner, Lagrange's method of envelopes, with similar results for the case of the eikonal equation.

In Section 2 we present an elementary constructive proof of the fact that any two complete solutions of the HJ equation are related by a time-independent canonical transformation and that any complete solution of the HJ equation is obtained from any other such solution looking for the envelope of a family of surfaces. In Section 3 we apply these results to geometrical optics.

\section{Complete solutions of the HJ equation, canonical transformations, and envelopes}
For a given Hamiltonian, $H(q_{i}, p_{i}, t)$, of a system with $n$ degrees of freedom, the HJ equation is the first-order PDE
\begin{equation}
H \left( q_{i}, \frac{\partial S}{\partial q_{i}}, t \right) + \frac{\partial S}{\partial t} = 0. \label{hje}
\end{equation}
A complete solution of this equation is a function of $2n + 1$ variables, $S(q_{i}, P_{i}, t)$, that satisfies Eq.\ (\ref{hje}) and the condition
\[
\det \left( \frac{\partial^{2} S}{\partial q_{i} \partial P_{j}} \right) \not= 0.
\]
The function $S$ generates a canonical transformation
\begin{equation}
Q_{i} = Q_{i}(q_{j}, p_{j}, t), \qquad P_{i} = P_{i}(q_{j}, p_{j}, t), \label{ct}
\end{equation}
such that the new Hamiltonian is equal to zero (see, e.g., Ref.\ \cite{Ar}). The canonical transformation (\ref{ct}) is determined by
\begin{equation}
p_{i} = \frac{\partial S}{\partial q_{i}}, \qquad Q_{i} = \frac{\partial S}{\partial P_{i}}, \label{par}
\end{equation}
which give $p_{i}$ and $Q_{i}$ as functions of $q_{i}$, $P_{i}$, and $t$, hence,
\begin{equation}
{\rm d} S = p_{i}(q_{j}, P_{j}, t) \, {\rm d} q_{i} - H \big( q_{i}, p_{i}(q_{j}, P_{j}, t), t \big) \, {\rm d} t + Q_{i}(q_{j}, P_{j}, t) \, {\rm d} P_{i} \label{difs}
\end{equation}
(here and henceforth there is sum over repeated indices).

In a similar manner, if $\tilde{S}(q_{i}, \tilde{P}_{i}, t)$ is a second complete solution of the HJ equation (\ref{hje}), then
\begin{equation}
{\rm d} \tilde{S} = p_{i}(q_{j}, \tilde{P}_{j}, t) \, {\rm d} q_{i} - H \big( q_{i}, p_{i}(q_{j}, \tilde{P}_{j}, t), t \big) \, {\rm d} t + \tilde{Q}_{i}(q_{j}, \tilde{P}_{j}, t) \, {\rm d} \tilde{P}_{i}, \label{difst}
\end{equation}
where $(\tilde{Q}_{i}, \tilde{P}_{i})$ is another set of canonical coordinates, all of which are constants of motion. By combining Eqs.\ (\ref{difs}) and (\ref{difst}) one finds that
\begin{equation}
\tilde{Q}_{i} {\rm d} \tilde{P}_{i} - Q_{i} {\rm d} P_{i} = {\rm d} (\tilde{S} - S), \label{gf}
\end{equation}
{\em provided that}\/ $P_{i}$ and $\tilde{P}_{i}$ are related in such a way that
\begin{equation}
\frac{\partial (\tilde{S} - S)}{\partial q_{i}} = 0 \label{conde}
\end{equation}
[see the first equation in (\ref{par})]. The condition
\[
\frac{\partial (\tilde{S} - S)}{\partial t} = 0
\]
is automatically satisfied as a consequence of Eqs.\ (\ref{conde}) since, by hypothesis, $S$ and $\tilde{S}$ are solutions of the same HJ equation (see Example 1, below). Equations (\ref{conde}) constitute a system of $n$ equations that, under appropriate conditions, determine the $q_{i}$ as functions of $P_{i}$, $\tilde{P}_{i}$, and $t$.

Equation (\ref{gf}) shows that the coordinates $(Q_{i}, P_{i})$ and $(\tilde{Q}_{i}, \tilde{P}_{i})$, associated with $S$ and $\tilde{S}$, respectively, are related by a {\em time-independent canonical transformation}, and that
\begin{equation}
F \equiv \tilde{S}(q_{i}, \tilde{P}_{i}, t) - S(q_{i}, P_{i}, t) \label{gff}
\end{equation}
is a generating function of this transformation {\em if}\/ $(P_{i}, \tilde{P}_{i})$ are functionally independent. In this latter case, from Eq.\ (\ref{gf}), we find that
\begin{equation}
\tilde{Q}_{i} = \frac{\partial F}{\partial \tilde{P}_{i}}, \qquad Q_{i} = - \frac{\partial F}{\partial P_{i}}. \label{gf2}
\end{equation}

Thus, for a given HJ equation, the only difference between its complete solutions is in the sets of canonical coordinates $(Q_{i}, P_{i})$ associated with them. These coordinates are constants of motion that label the solutions of the equations of motion.

\vspace*{1ex}

\noindent {\bf Remark 1.} Even if $(P_{i}, \tilde{P}_{i})$ are functionally dependent, Eq.\ (\ref{gf}) implies that $F$ can be expressed as a function of $P_{i}$ and $\tilde{P}_{i}$ only, though not in a unique way. As a consequence of Eqs.\ (\ref{conde}), $F(P_{i}, \tilde{P}_{i}) = 0$ is the (equation of the) {\em envelope}\/ of the family of surfaces $\tilde{S}(q_{i}, \tilde{P}_{i}, t) - S(q_{i}, P_{i}, t) = 0$, parameterized by the $q_{i}$.

\vspace*{1ex}

\noindent {\bf Example 1.} {\it Relation between two given complete solutions of the HJ equation}

By means of direct computations one can verify that the functions
\begin{equation}
S(q, P, t) = - mgtq + Pq - \frac{P^{2}t}{2m} + \frac{Pgt^{2}}{2} - \frac{1}{6} mg^{2}t^{3} \label{s1}
\end{equation}
and
\begin{equation}
\tilde{S}(q, \tilde{P}, t) = - mgtq + \frac{m}{2t} \left( q + \frac{gt^{2}}{2} + \tilde{P} \right)^{2} - \frac{1}{6} m g^{2} t^{3} \label{s2}
\end{equation}
are complete solutions of the HJ equation
\begin{equation}
\frac{1}{2m} \left( \frac{\partial S}{\partial q} \right)^{2} + mgq + \frac{\partial S}{\partial t} = 0, \label{hjeje}
\end{equation}
corresponding to a particle of mass $m$ in a uniform gravitational field. Then,
\begin{equation}
\tilde{S} - S = \frac{m}{2t} \left( q + \frac{gt^{2}}{2} + \tilde{P} \right)^{2} - Pq + \frac{P^{2}t}{2m} - \frac{Pgt^{2}}{2} \label{difs2}
\end{equation}
and Eq.\ (\ref{conde}) yields
\begin{equation}
\frac{m}{t} \left( q + \frac{gt^{2}}{2} + \tilde{P} \right) - P = 0. \label{relq}
\end{equation}
Making use of this last relation in order to eliminate $q$ from Eq.\ (\ref{difs2}) we obtain [see Eq.\ (\ref{gff})]
\begin{equation}
F(P, \tilde{P}) = P \tilde{P}. \label{fgcs}
\end{equation}
It may be noticed that $t$ disappeared together with $q$, and that the canonical transformation generated by the generating function (\ref{fgcs}) is the ``exchange transformation,'' $\tilde{Q} = P$, $\tilde{P} = - Q$ [see Eqs.\ (\ref{gf2})].

On the other hand, assuming that $(P_{i}, \tilde{P}_{i})$ are functionally independent, from Eq.\ (\ref{gff}) we have
\begin{equation}
\tilde{S}(q_{i}, \tilde{P}_{i}, t) = S(q_{i}, P_{i}, t) + F(P_{i}, \tilde{P}_{i}). \label{rel}
\end{equation}
In order to express $\tilde{S}$ as a function of $q_{i}$, $\tilde{P}_{i}$, and $t$ only, the parameters $P_{i}$ appearing on the right-hand side of this last equation are eliminated making use of the relations [see Eqs.\ (\ref{par}) and (\ref{gf2})]
\begin{equation}
\frac{\partial (S + F)}{\partial P_{i}} = 0, \label{condes}
\end{equation}
which are analogous to Eqs.\ (\ref{conde}). In this case, owing to Eqs.\ (\ref{condes}), $\tilde{S}(q_{i}, \tilde{P}_{i}, t) = 0$ is the envelope of the family of surfaces $S(q_{i}, P_{i}, t) + F(P_{i}, \tilde{P}_{i}) = 0$, parameterized by the $P_{i}$.

The remaining case, where $(P_{i}, \tilde{P}_{i})$ are functionally dependent, can be reduced to the first case by replacing one or several parameters $P_{i}$, contained in $S$, by their conjugate variables, $Q_{i}$, in order to obtain a set of $n$ parameters that together with the $\tilde{P}_{i}$ form a functionally independent set. For instance, instead of $S(q_{i}, P_{i}, t)$ we can make use of its Legendre transform $S'(q_{i}, Q_{i}, t) \equiv S(q_{i}, P_{i}, t) - P_{i} Q_{i}$, if $(Q_{i}, \tilde{P}_{i})$ are functionally independent. Then, by relabeling the parameters in $S'$, we obtain a function $S'(q_{i}, P_{i}, t)$ that will be related to $\tilde{S}(q_{i}, \tilde{P}_{i}, t)$ according to (\ref{rel}) and (\ref{condes}).

Thus, we conclude that from a given complete solution of the HJ equation, or a Legendre transform of it, one can obtain any other complete solution by means of an appropriate canonical transformation.

\vspace*{1ex}

\noindent {\bf Example 2.} {\it Obtaining a complete solution of the HJ equation from another complete solution}

It is illustrative to consider again the complete solution (\ref{s1}) of the HJ equation (\ref{hjeje}), with $F$ given by Eq.\ (\ref{fgcs}), so that the right-hand side of Eq.\ (\ref{rel}) is
\begin{equation}
S(q, P, t) + F(P, \tilde{P}) = - mgtq + Pq - \frac{P^{2}t}{2m} + \frac{Pgt^{2}}{2} - \frac{1}{6} mg^{2}t^{3} + P \tilde{P}. \label{nn}
\end{equation}
Then, Eq.\ (\ref{condes}) yields
\[
q - \frac{Pt}{m} + \frac{gt^{2}}{2} + \tilde{P} = 0,
\]
which is equivalent to Eq.\ (\ref{relq}). Making use of this last relation to eliminate $P$ from the right-hand side of Eq.\ (\ref{nn}), one recovers the complete solution (\ref{s2}) of Eq.\ (\ref{hjeje}).

\vspace*{1ex}

\noindent {\bf Remark 2.} If the number of parameters $\tilde{P}_{i}$ contained in $F$ is less than $n$, the solution $\tilde{S}$ of the HJ equation obtained by means of Eqs.\ (\ref{rel}) and (\ref{condes}) will not be complete but, nevertheless, it will be a solution. Indeed, we can think that $F$ originally had $n$ parameters $\tilde{P}_{i}$, and that $k$ of them have taken some fixed values, leaving only $n - k$ arbitrary parameters $\tilde{P}_{i}$.

\section{Geometrical optics}
As pointed out in the Introduction, in classical mechanics, the knowledge of a single complete solution of the HJ equation is enough to find the solution of the equations of motion, so that usually there would be no need of additional complete solutions of the HJ equation. By contrast, in geometrical optics it is important to have different (not necessarily complete) solutions of the eikonal equation, which is analogous to the HJ equation.

The eikonal equation is given by
\begin{equation}
(\nabla S)^{2} = n^{2}, \label{eikeq}
\end{equation}
where $n$ is the refractive index of the medium, and $S$ and $n$ are functions of the space coordinates only (assuming that the medium is isotropic) (see, e.g., Refs.\ \cite{Sy} and \cite{BW}). In terms of Cartesian coordinates, $(x, y, z)$, the eikonal equation (\ref{eikeq}) can be written, e.g., as
\begin{equation}
\pm \left[ n^{2} - \left( \frac{\partial S}{\partial x} \right)^{2} - \left( \frac{\partial S}{\partial y} \right)^{2} \right]^{1/2} + \frac{\partial S}{\partial z} = 0, \label{hjeik}
\end{equation}
which has the form of the HJ equation (\ref{hje}), with $z$ in place of $t$. (See also Ref.\ \cite{KW}.)

The surfaces $S = {\rm const.}$ are the wavefronts and the rays of light are the curves orthogonal to the wavefronts. Each solution of the eikonal equation corresponds to a family of wavefronts in such a way that, if $a_{1}$ and $a_{2}$ are constants (such that $S = a_{1}$ and $S = a_{2}$ are nonempty sets), the wavefront $S = a_{2}$ is obtained from $S = a_{1}$ by the propagation of the light by a time $(a_{2} - a_{1})/c$, where $c$ is the speed of light in vacuum.

For instance, if the refractive index is {\em constant}, one can readily verify that
\begin{equation}
S(x, y, P_{1}, P_{2}, z) = n \sqrt{(x - P_{1})^{2} + (y - P_{2})^{2} + z^{2}} \label{sw}
\end{equation}
is a complete solution of the eikonal equation [with $P_{1}, P_{2} \in  (- \infty, \infty)$] and the wavefronts $S = {\rm const.}$ are spheres [centered at $(P_{1}, P_{2}, 0)$]. Choosing the generating function
\begin{equation}
F(P_{i}, \tilde{P}_{i}) = P_{i} \tilde{P}_{i} \label{exch}
\end{equation}
[cf.\ Eq.\ (\ref{fgcs})], which generates the exchange transformation, $\tilde{Q}_{i} = P_{i}$, $\tilde{P}_{i} = - Q_{i}$, from Eqs.\ (\ref{condes}) we obtain the conditions
\[
- \frac{n (x - P_{1})}{\sqrt{(x - P_{1})^{2} + (y - P_{2})^{2} + z^{2}}} + \tilde{P}_{1} = 0, \qquad - \frac{n (y - P_{2})}{\sqrt{(x - P_{1})^{2} + (y - P_{2})^{2} + z^{2}}} + \tilde{P}_{2} = 0.
\]
Making use of these relations to eliminate the $P_{i}$ from $S + F$ [see Eq.\ (\ref{rel})], we obtain a second complete solution of the eikonal equation, namely
\begin{equation}
\tilde{S}(x, y, \tilde{P}_{1}, \tilde{P}_{2}, z) = \tilde{P}_{1} x + \tilde{P}_{2} y + \sqrt{n^{2} - \tilde{P}_{1}{}^{2} - \tilde{P}_{2}{}^{2}} \, z \label{pw}
\end{equation}
(with $\tilde{P}_{1}{}^{2} + \tilde{P}_{2}{}^{2} \leqslant n^{2}$). The wavefronts $\tilde{S} = {\rm const.}$ are planes [with normal $(\tilde{P}_{1}, \tilde{P}_{2}, \sqrt{n^{2} - \tilde{P}_{1}{}^{2} - \tilde{P}_{2}{}^{2}})$].

Other generating functions lead to other families of wavefronts. According to the results of Section 2, with a suitable function $F$, one can obtain any possible family of wavefronts, starting from {\em any}\/ complete solution of the eikonal equation [such as (\ref{sw}) or (\ref{pw})].

\vspace*{1ex}

\noindent {\bf Example 3.} {\it Refracted wavefronts produced by a point source}

In this example we shall consider two homogeneous media, of (constant) refractive indices $n_{0}$ and $n_{1}$, separated by a plane, and we want to find the wavefronts in the second medium produced by a point source in the first medium.

A complete solution of the eikonal equation in the second medium is given by
\begin{equation}
S(x, y, P_{1}, P_{2}, z) = P_{1} x + P_{2} y + \sqrt{n_{1}{}^{2} - P_{1}{}^{2} - P_{2}{}^{2}} \, z, \label{pw2}
\end{equation}
its wavefronts are planes and the parameters $P_{1}$, $P_{2}$ specify the normal to the planes and, therefore, the direction of the light rays. We want to replace the parameters $P_{1}, P_{2}$ by some new parameters, $\tilde{P}_{1}, \tilde{P}_{2}$, that specify the position of the point source in the first medium. To this end we find the light rays emanating from a point source placed at $(x_{0}, y_{0}, z_{0})$. Assuming that
\[
n(x, y, z) = \left\{ \begin{array}{ll} n_{0}, & {\rm if\ } z < 0, \\ n_{1}, & {\rm if\ } z > 0, \end{array} \right.
\]
from the Hamilton equations applied to the Hamiltonian
\[
H = - \sqrt{n^{2} - p_{x}{}^{2} - p_{y}{}^{2}}
\]
[see Eq.\ (\ref{hjeik})], we find that, since $H$ does not depend on $x$ or $y$, $p_{x}$ and $p_{y}$ are constant (which is equivalent to Snell's law) and
\begin{equation}
\frac{{\rm d} x}{{\rm d} z} = \frac{\partial H}{\partial p_{x}} = \frac{p_{x}}{\sqrt{n^{2} - p_{x}{}^{2} - p_{y}{}^{2}}}, \qquad \frac{{\rm d} y}{{\rm d} z} = \frac{\partial H}{\partial p_{y}} = \frac{p_{y}}{\sqrt{n^{2} - p_{x}{}^{2} - p_{y}{}^{2}}}. \label{rec}
\end{equation}
These last two equations mean that the light rays are straight lines for $z < 0$, or $z > 0$ (where $n$ is constant). Integrating both sides of the first equation in (\ref{rec}) over $z$, between $z_{0}$ and 0 (with $z_{0} < 0$), we obtain
\begin{equation}
x_{1} - x_{0} = - \frac{z_{0} p_{x}}{\sqrt{n_{0}{}^{2} - p_{x}{}^{2} - p_{y}{}^{2}}}, \label{ex}
\end{equation}
where $x_{0}$ and $x_{1}$ are the values of $x$ at $z = z_{0}$ and $z = 0$, respectively
and, in a similar manner,
\begin{equation}
y_{1} - y_{0} = - \frac{z_{0} p_{y}}{\sqrt{n_{0}{}^{2} - p_{x}{}^{2} - p_{y}{}^{2}}}. \label{ye}
\end{equation}

Being the solution of the Hamilton equations, the relation between the values of the canonical coordinates $(x, y, p_{x}, p_{y})$ at $z = z_{0}$ and at $z = 0$ must be a canonical transformation (in fact, we shall find its generating function below). Since we are looking for a canonical transformation to replace the parameters $P_{i}$ that specify the direction of the light rays in the second medium by parameters $\tilde{P}_{i}$ that specify the position of the point source, the values of the canonical coordinates $(x, y, p_{x}, p_{y})$ at $z = 0$ will be denoted as $(Q_{1}, Q_{2}, P_{1}, P_{2})$, and those at $z = z_{0}$ will be denoted as $(\tilde{P}_{1}, \tilde{P}_{2}, - \tilde{Q}_{1}, - \tilde{Q}_{2})$ (the minus signs are necessary to have a canonical transformation); then, from Eqs.\ (\ref{ex}) and (\ref{ye}), we obtain
\begin{eqnarray*}
\tilde{Q}_{1} & = & - P_{1}, \qquad \tilde{P}_{1} = Q_{1} + \frac{z_{0} P_{1}}{\sqrt{n_{0}{}^{2} - P_{1}{}^{2} - P_{2}{}^{2}}}, \\
\tilde{Q}_{2} & = & - P_{2}, \qquad \tilde{P}_{2} = Q_{2} + \frac{z_{0} P_{2}}{\sqrt{n_{0}{}^{2} - P_{1}{}^{2} - P_{2}{}^{2}}}.
\end{eqnarray*}
A straightforward computation shows that $\tilde{Q}_{i} {\rm d} \tilde{P}_{i} - Q_{i} {\rm d} P_{i}$ is the differential of [see Eq.\ (\ref{gf})]
\begin{equation}
F(P_{i}, \tilde{P}_{i}) = - P_{1} \tilde{P}_{1} - P_{2} \tilde{P}_{2} - z_{0} \sqrt{n_{0}{}^{2} - P_{1}{}^{2} - P_{2}{}^{2}}. \label{W}
\end{equation}
(Note that $z_{0}$ is treated as a parameter.) The generating function (\ref{W}) coincides with the {\em characteristic function}\/ $W$ obtained by calculating optical lengths in Ref.\ \cite{Sy} [Eq.\ (6$\cdot$23)].

Then, from Eqs.\ (\ref{rel}), (\ref{pw2}), and (\ref{W}), we obtain
\begin{equation}
\tilde{S} = P_{1} x + P_{2} y + \sqrt{n_{1}{}^{2} - P_{1}{}^{2} - P_{2}{}^{2}} \, z - P_{1} \tilde{P}_{1} - P_{2} \tilde{P}_{2} - z_{0} \sqrt{n_{0}{}^{2} - P_{1}{}^{2} - P_{2}{}^{2}}, \label{refr}
\end{equation}
with the $P_{i}$ determined by the conditions [see Eqs.\ (\ref{condes})]
\begin{eqnarray}
x - \frac{z P_{1}}{\sqrt{n_{1}{}^{2} - P_{1}{}^{2} - P_{2}{}^{2}}} - \tilde{P}_{1} + \frac{z_{0} P_{1}}{\sqrt{n_{0}{}^{2} - P_{1}{}^{2} - P_{2}{}^{2}}} & = & 0, \label{refx} \\
 y- \frac{z P_{2}}{\sqrt{n_{1}{}^{2} - P_{1}{}^{2} - P_{2}{}^{2}}} - \tilde{P}_{2} + \frac{z_{0} P_{2}}{\sqrt{n_{0}{}^{2} - P_{1}{}^{2} - P_{2}{}^{2}}} & = & 0. \label{refy}
\end{eqnarray}
In this case, instead of giving the refracted wavefronts in the implicit form $\tilde{S} = {\rm const.}$, it is simpler to express them in parameterized form: for a given value of $\tilde{S}$, Eqs.\ (\ref{refr})--(\ref{refy}) can be employed to find the coordinates $x$, $y$, $z$ of the points of the wavefront in terms of the parameters $P_{1}$ and $P_{2}$ (with the coordinates of the source, $(\tilde{P}_{1}, \tilde{P}_{2}, z_{0}) = (x_{0}, y_{0}, z_{0})$, fixed).

It should be remarked that the eikonal functions (\ref{sw}), (\ref{pw}), (\ref{pw2}), and (\ref{refr}) correspond to homogeneous media (i.e., constant refractive indices), but that the basic results are applicable to any isotropic medium.

\section{Concluding remarks}
The results of Section 2 show that the Lagrange method of envelopes arises in a natural way in the cases considered here, and that the new solutions obtained in this manner have a very special structure, being sums of the form (\ref{rel}). Furthermore, we have shown that the function $F$, appearing in Eq.\ (\ref{rel}), is related to a canonical transformation. It may be pointed out that the results derived here are applicable to any first-order PDE that contains the unknown function $S$ only through its derivatives.

As we have shown in Section 3, the formalism of Hamiltonian mechanics can be conveniently applied to the geometrical optics, allowing us to derive the relevant relations without having to use other standard tools (such as the Fermat principle, Snell's law, or the concept of optical length).

\section*{Acknowledgements}
The authors wish to thank Gilberto Silva Ortigoza and Ira\'is Rubalcava Garc\'ia for helpful discussions. One of the authors (G.S.A.G.) also wishes to thank the Vicerrector\'{\i}a de Investigaci\'on y Estudios de Posgrado of the Universidad Aut\'onoma de Puebla for financial support.

\end{document}